\documentclass[conference]{IEEEtran}
\IEEEoverridecommandlockouts
\usepackage{cite}
\usepackage{amsmath,amssymb,amsfonts}
\usepackage{algorithm}
\usepackage{graphicx}
\usepackage{textcomp}
\usepackage{xcolor}
\usepackage{times}
\usepackage{multicol}
\usepackage{pgfplots}
\usepackage{code}
\usepackage{lscape}
\usepackage{fancybox}
\usepackage{algpseudocode}
\usepackage{listings}
\usepackage{comment}
\usepackage{url}

\def\BibTeX{{\rm B\kern-.05em{\sc i\kern-.025em b}\kern-.08em
    T\kern-.1667em\lower.7ex\hbox{E}\kern-.125emX}}

\lstdefinestyle{BashInputStyle}{
  language=java,
  basicstyle=\small\sffamily,
  numbers=left,
  numberstyle=\tiny,
  numbersep=3pt,
  frame=tb,
  linewidth=0.9\linewidth,
  xleftmargin=0.1\linewidth
}

\definecolor{dkgreen}{rgb}{0,0.6,0}
\definecolor{gray}{rgb}{0.5,0.5,0.5}
\definecolor{mauve}{rgb}{0.58,0,0.82}

\begin{document}

\title{Evaluating Developer-written Unit Test Case Reduction for Java - A Replication Study}

\author{
\IEEEauthorblockN{Tuan D Le}
\IEEEauthorblockA{\textit{School of Computing} \\
\textit{Weber State University}\\
Ogden, UT, USA \\
tuanle@mail.weber.edu }
\and
\IEEEauthorblockN{Brandon Wilber}
\IEEEauthorblockA{\textit{School of Computing} \\
\textit{Weber State University}\\
Ogden, UT, USA \\
brandon.s.wilber@gmail.com}
\and

\IEEEauthorblockN{Arpit Christi}
\IEEEauthorblockA{\textit{School of Computing} \\
\textit{Weber State University}\\
Ogden, UT, USA \\
arpitchristi@weber.edu}

}

\maketitle

\begin{abstract}
Failing test case reduction can promote efficient debugging because a developer may not need to observe components that are not relevant to inducing failure. Failing test case reduction can also improve efficient of fault localization. All of these have prompted researchers to study the reduction process, the reduction output, and the removed entities. Christi et al. studied test reduction using a tool called ReduSharptor for C\# tests. They considered the test to be an Abstract Syntax Tree (AST). Based on that, they studied the reduction outcome and removed entities in terms of Leaf nodes and Non-Leaf nodes of the AST. They claimed that (1) leaf nodes are removed in large numbers and (2) probability of removal is slightly higher than non-leaf nodes. 

We replicate their results using a different test case reduction tool ReduJavator for Java unit tests. We evaluate test reduction using 30 randomly chosen bugs from \textit{Defects4J} database and 30 mutants for 6 open source projects. Our results confirm their first claim: leaf nodes are removed in large numbers. Our results are inconclusive to support their second claim; we cannot confirm that probability of removals is higher for non-leaf nodes.

\end{abstract}

\begin{IEEEkeywords}
program debugging; software testing; test case reduction.
\end{IEEEkeywords}

\section{Introduction}

Debugging is challenging, time-consuming, and difficult even if the developer has a failing test case at their disposal. Test case reduction can help the developer by removing non-failure-inducing components from the test. As developer needs to observe fewer non-failure-inducing components, they will be able to reach the failure-inducing component faster. Apart from being helpful to the developer, test reduction also improves fault-localization(FL)\cite{christi18reduce,vince2021reduction}. 

Weber and Christi proposed a tool ReduSharptor to reduce C\# unit tests~\cite{weber2024redusharptor}. ReduSharptor uses Abstract Syntax Tree (AST) presentation of the unit test to reduce the test. Later they studied reduction output (the reduced test) and the removed entities in terms of its location in AST (the AST is generated using Roslyn C\# compiler\cite{roslyn-documentation})~\cite{Christi2024OnReducibility}. They divided the statement nodes in AST into tree-statement and non-tree statement. Non-tree statements are leaf nodes while tree-statement are non-leaf nodes that has leaf-nodes further in the tree. They reduced tests for 30 real-world failing unit tests to study the reduction process, outcome and removed entities. Based on that they claim that (1) On average, non-tree-statements are removed 50 times higher than tree statements. (2) On average, probability of removal is 1.7 times higher for non-tree-statements. 

Brandon et al. proposed a tool ReduJavator to reduce Java unit tests~\cite{Wilber2024ReduJavator}. To replicate the study performed by Christi and Weber, we perform the exact same experiments, using the same process and measurements while using Java subjects and tests. As they use 30 real-world bugs from open source projects, we also use 30 real-world bugs taken from \textit{Defects4J} bug database and 30 mutants taken from 6 open-source java projects~\cite{just2014defects4j}. We compare our results with their results to confirm or reject their claims. 

\subsection{Motivation}
In their work titled  ``Synthesizing Input Grammars: A Replication Study'', Bendrissou et al. lamented the fact that replication studies are rare in software engineering~\cite{bendrissou2022synthesizing}. They challenged other researchers to conduct more replication study to reduce threats to validity. They also exhort to make data, tools, experiments and results available for further study and comparison. They want all these to be norm in software engineering and not an exception as such replicability is common in other science fields. Recently, we have seen many conferences in Softwaren Engineering have ``Negative Results or Reproducibility Study'' tracks for the same purposes.  

We agree with Bendrissou et al. on the need of replication studies. We replicate study conducted by previous researchers using different tool and different subjects. We even add type of bugs: The original research just study real-world bugs while we study both real-world bugs and mutants. The purpose of our study is to accept or challenge the claims by previous researchers based on the same experiments conducted on different subjects using different tool.

\subsection{Contribution}
Our contribution is as follows. 
\begin{enumerate}
    \item We conduct a replication study of previous research using same experiments, process and measurements but using different subjects and tools. 
    \item Based on our results we support their claim that non-tree-statements are reduced in large numbers than tree-statements. 
    \item Based on our results we challenge their claim that non-tree-statements have slightly higher probability of removal than tree-statements. 
\end{enumerate}

The ReduSharptor tool for C\# test reduction is available at https://github.com/amchristi/ReduSharptor. The ReduJavator tool for Java test reduction is available at https://github.com/amchristi/ReduJavator.

\section{Related Work}\label{related-work}
Delta Debugging (DD) algorithm was proposed to simpify and reduce failure-inducing tests while keeping the failure. Hierarchical Delta Debugging (HDD) algorithm is an extension of DD algorithm that works better on test inputs that are hierarchical like HTML, programs, JSON, XML etc. Many algorithms, techniques and tools have been proposed to reduce different kind of failing test inputs. Some of the recent techniques include but not limited to ReduJavator, ReduSharptor, ReduKtor, ProdDD, Picirency, Perses and DDSET~\cite{Wilber2024ReduJavator, weber2024redusharptor, stepanov2019reduktor, wang2023probabilistic, want2021probabilistic, hodovan2016modernizing, perses, gopinath2020abstracting}.

Christi and Weber studied test reduction outcome and the removed entities in terms of the location of test program statements within the test~\cite{Christi2024OnReducibility}. In their ``threats to validity'' section, they mentioned that their results are only for C\# subjects and speculated that the results will generalize for other programming languages. 

Bendrissou et al. conducted a replication study on Synthesizing Input Grammar by implementing an existing algorithm using a new tool and utilizing the tool on different subjects. They noticed that their results don't confirm the claims made by previous researchers. They implored the need for more replication studies and reproducibility in Software Engineering research.

\section{Experiments}\label{experiments}
In this section, we describe the experiments that we perform. 
\subsection{Definitions and Terminology from Pervious Work}
We aim to replicate the study conducted in previous work by Christi and Weber. The following terms and definitions have already been discussed in their work. We enumerate them here as we will use them for the remaining of the paper. 
\begin{enumerate}
    \item \textit{original-test}
    \item \textit{minimal-test}
    \item \textit{reduced-entities}
    \item \textit{TreeNode/TreeStmt}
    \item \textit{NonTreeNode/NonTreeStmt}
    
\end{enumerate}
Please refer to the original work for complete definitions~\cite{Christi2024OnReducibility}.  

We want to replicate the study on impact of statement categories on the removal process and the reduction outcome. For that, we use the subjects and procedures establish in the previous research work by Wilber et al. to evaluate their ReduJavator tool for accuracy and applicability~\cite{Wilber2024ReduJavator}.

\subsection{Subjects}
Wilber et al. utilized 30 real-world bugs and failing tests from the \textit{Defects4J} bug database for their evaluations. They also utilized 30 mutants that resulted in single failing test each for six open source Java projects. For the details of the bugs, please refer to Table I and Table II of their paper. We utilized the exact same bugs for our evaluations. We don't create 30 new mutants to evaluate. We continue to utilize the exact same mutants that they used.

\begin{table*}
\caption{Mutants Test, Project, Total Stmts, \textit{\#NTN} - Number of \textit{NonTreeNodes}, \textit{\#TN} - Number of \textit{TreeNodes}, ARS, PRS, ANTRS, PNTRS, ATRS, PTRS }
\begin{center}
{\scriptsize
\begin{tabular}{|l|l| r|r|r|r|r|r|r|r|r|}
\hline
Test & Project & Stmts &  \textit{\#NTN}  & \textit{\#TN} & ARS & PRS &  ANTRS & PNTRS & ATRS & PTRS   \\
\hline
\hline
{ testOnlyReturnErrors} & commons-validator  & 13 & 13 & 0 & 7 & 53.84\% & 7 & 53.84\% & 0 & 0\%\\
\hline
{ testValidatorException} & commons-validator  & 7 & 6 & 1 & 3 & 42.85\% & 3 & 42.85\% & 0 & 0\%\\
\hline
{ testManualBooleanDeprecated} & commons-validator  & 31 & 29 & 2 & 27 & 87.09\% & 25 & 80.64\% & 2 & 6.45\%\\
\hline
{ testManualBoolean} & commons-validator  & 31 & 29 & 2 & 27 & 87.09\% & 25 & 80.65\% & 2 & 6.45\%\\
\hline
{ checkMissingMethods} &  commons-math &  10 & 8 & 2 & 3 & 30.00\% & 3 & 30.00\% & 0 & 0\%\\
\hline
{ testConvergenceOnFunctionAccuracy} &  commons-math &  10 & 10 & 0 & 4 & 40.00\% & 4 & 40.00\% & 0 & 0\%\\
\hline
{ testQuadraticFunction} &  commons-math &  20 & 20 & 0 & 5 & 25.00\% & 5 & 25.00\% & 0 & 0\%\\
\hline
{ testParameters} &  commons-math &  9 & 7 & 2 & 2 & 22.22\% & 2 & 22.22\% & 0 & 0\%\\
\hline
{ testQuinticFunction} &  commons-math &  25 & 25 & 0 & 11 & 44.00\% & 11 & 44.00\% & 0 & 0\%\\
\hline
{ ExtendedBufferedReaderTest} & commons-csv &  19 & 13 & 6 & 11 & 57.89\% & 6 & 31.57\% & 5 & 26.31\%\\
\hline
{ testReadLine} & commons-csv &  34 & 29 & 5 & 29 & 85.29\% & 25 & 73.52\% & 4 & 11.76\%\\
\hline
{ testSerialization} & commons-csv &  25 & 21 & 4 & 20 & 80.00\% & 18 & 72.00\% & 2 & 8.00\%\\
\hline
{ testJdbcPrinterWithResultSetHeader} & commons-csv &  11 & 7 & 4 & 3 & 27.27\% & 3 & 27.27\% & 0 & 0\%\\
\hline
{ testPrintRecordsWithEmptyVector} & commons-csv &  10 & 8 & 2 & 10 & 100\% & 8 & 80\% & 2 & 20\%\\
\hline
{ testLRUBehavior} & commons-dbcp &  28 & 26 & 2 & 5 & 17.85\% & 5 & 17.85\% & 0 & 0\%\\
\hline
{ testPreparedStatementPooling} & commons-dbcp &  17 & 15 & 2 & 8 & 47.05\% & 7 & 41.17\% & 1 & 5.88\%\\
\hline
{ testPStmtPoolingWithNoClose} & commons-dbcp &  20 & 20 & 0 & 4 & 20\% & 4 & 20\% & 0 & 0\%\\
\hline
{ testCallableStatementPooling} & commons-dbcp &  34 & 33 & 1 & 9 & 26.47\% & 9 & 26.47\% & 0 & 0\%\\
\hline
{ testClosePool} & commons-dbcp &  18 & 16 & 2 & 10 & 55.55\% & 8 & 44.44\% & 1 & 5.55\%\\
\hline
{ testMySqlNullOutput} & commons-csv& 81  & 72 & 9 & 75 & 92.59\% & 67 & 82.71\% & 8 & 9.87\%\\
\hline
{ testExcelHeaderCountLessThanData} & commons-csv &  8 & 7 & 1 & 6 & 75\% & 5 & 62.5\% & 1 & 12.5\%\\
\hline
{ testToMapWithNoHeader} & commons-csv &  5 & 5 & 0 & 2 & 40\% & 2 & 40\% & 0 & 0\%\\
\hline
{ testManualObject} & commons-validator &  22 & 20 & 2 & 22 & 100.00\% & 20 & 90.90\% & 2 & 9.09\%\\
\hline
{ testValidator288} & commons-validator &  11 & 11 & 0 & 7 & 63.63\% & 7 & 63.63\% & 0 & 0\%\\
\hline
{ testOptionalFinish} & commons-compress &  11 & 9 & 2 & 2 & 18.18\% & 1 & 9.09\% & 1 & 9.09\%\\
\hline
{ testFileEntryFromPath} & commons-compress &  18 & 15 & 3 & 8 & 44.44\% & 6 & 33.33 \% & 2 & 11.11\%\\
\hline
{ testExplicitFileEntry} & commons-compress& 21 & 18 & 3 & 11 & 52.38\% & 9 & 42.85\% & 2 & 9.52\%\\
\hline
{ testTarArchiveLongNameCreation} & commons-compress & 31 & 29 & 2 & 16 & 51.61\% & 15 & 48.38\% & 1 & 3.22\%\\
\hline
{ testContent} & commons-compress & 15 & 13 & 2 & 7 & 46.66\% & 6 & 40\% & 1 & 6.66\%\\
\hline
{ testReparse} & commons-compress & 32 & 32 & 0 & 20 & 62.5\% & 20 & 62.5\% & 0 & 0\%\\
\hline
\hline
{ \textbf{Mean} } &  & \textbf{20.9} & \textbf{18.86} & \textbf{2.03} & \textbf{12.46} & \textbf{53.21\%} & \textbf{11.2} & \textbf{47.64\%} & \textbf{1.23}  & \textbf{5.38\%}\\
\hline

\end{tabular}
}
\end{center}
\label{tab:results}
\end{table*}

\begin{table*}
\caption{\textit{Defects4J} Test, Project, Total Stmts, \textit{\#NTN} - Number of \textit{NonTreeNodes}, \textit{\#TN} - Number of \textit{TreeNodes}, ARS, PRS, ANTRS, PNTRS, ATRS, PTRS }
\begin{center}
{\scriptsize
\begin{tabular}{|l|l| r|r|r|r|r|r|r|r|r|}
\hline
Test & Project & Stmts &  \textit{\#NTN}  & \textit{\#TN} & ARS & PRS &  ANTRS & PNTRS & ATRS & PTRS   \\
\hline
\hline
TestLang747 & commons-lang& 26 & 26 & 0 & 25 & 96.15\% & 25 & 96.15\% & 0 & 0.00\% \\
\hline
testCreateNumber & commons-lang & 35 & 33 & 2 & 33 & 94.29\% & 31 & 88.57\% & 2 & 5.71\% \\
\hline
testCalendarTimezoneRespected & commons-lang & 14 & 12 & 2 & 12 & 85.71\% & 10 & 71.43\% & 2 & 14.29\% \\
\hline
testLANG807 & commons-lang & 6 & 5 & 1 & 2 & 33.33\% & 2 & 33.33\% & 0 & 0.00\% \\
\hline
testExceptions & commons-lang & 28 & 19 & 9 & 26 & 92.86\% & 18 & 64.29\% & 8 & 28.57\% \\
\hline
testConstructorEx7-TypeArray-intArray & joda-time & 31 & 25 & 6 & 28 & 90.32\% & 23 & 74.19\% & 5 & 16.13\% \\
\hline
testNormalizedStandard-periodType-months1 & joda-time & 13 & 13 & 0 & 6 & 46.15\% & 6 & 46.15\% & 0 & 0.00\% \\
\hline
testParseInto-monthDay-feb29-newYork-startOfYear & joda-time & 8 & 8 & 0 & 4 & 50.00\% & 4 & 50.00\% & 0 & 0.00\% \\
\hline
testForOffsetHoursMinutes-int-int & joda-time & 26 & 21 & 5 & 25 & 96.15\% & 20 & 76.92\% & 5 & 19.23\% \\
\hline
testForOffsetHoursMinutes-int-int & joda-time & 24 & 18 & 6 & 22 & 91.67\% & 17 & 70.83\% & 5 & 20.83\% \\
\hline
testLocaleIndependence & commons-codec & 17 & 12 & 5 & 1 & 5.88\% & 1 & 5.88\% & 0 & 0.00\% \\
\hline
testCodec98NPE & commons-codec & 16 & 16 & 0 & 8 & 50.00\% & 8 & 50.00\% & 0 & 0.00\% \\
\hline
testCodec101 & commons-codec & 8 & 8 & 0 & 2 & 25.00\% & 2 & 25.00\% & 0 & 0.00\% \\
\hline
testRfc4648Section10Encode & commons-codec & 21 & 21 & 0 & 18 & 85.71\% & 18 & 85.71\% & 0 & 0.00\% \\
\hline
testSoftLineBreakDecode & commons-codec & 38 & 38 & 0 & 19 & 50.00\% & 19 & 50.00\% & 0 & 0.00\% \\
\hline
testCpioUnarchive & commons-compress & 32 & 30 & 2 & 29 & 90.62\% & 27 & 84.38\% & 2 & 6.25\% \\
\hline
testArDelete & commons-compress & 54 & 49 & 5 & 52 & 96.30\% & 47 & 87.04\% & 5 & 9.26\% \\
\hline
testFinish & commons-compress & 26 & 21 & 5 & 23 & 88.46\% & 18 & 69.23\% & 5 & 19.23\% \\
\hline
testRead7ZipMultiVolumeArchiveForStream & commons-compress & 20 & 15 & 6 & 15 & 75.00\% & 10 & 50.00\% & 5 & 25.00\% \\
\hline
testParseOctalInvalid & commons-compress & 32 & 24 & 8 & 29 & 90.62\% & 22 & 68.75\% & 7 & 21.88\% \\
\hline
testGetLineNumberWithCR & commons-csv & 9 & 9 & 0 & 7 & 77.78\% & 7 & 77.78\% & 0 & 0.00\% \\
\hline
testToMapWithNoHeader & commons-csv & 5 & 5 & 0 & 2 & 40.00\% & 2 & 40.00\% & 0 & 0.00\% \\
\hline
testExcelHeaderCountLessThanData & commons-csv & 8 & 6 & 2 & 6 & 75.00\% & 4 & 50.00\% & 2 & 25.00\% \\
\hline
escape-caseSensitive & jsoup & 11 & 11 & 0 & 6 & 54.55\% & 6 & 54.55\% & 0 & 0.00\% \\
\hline
parsesQuiteRoughAttributes & jsoup & 7 & 7 & 0 & 5 & 71.43\% & 5 & 71.43\% & 0 & 0.00\% \\
\hline
testPseudoHas & jsoup & 36 & 36 & 0 & 27 & 75.00\% & 27 & 75.00\% & 0 & 0.00\% \\
\hline
handlesAbsPrefix & jsoup & 28 & 28 & 0 & 16 & 57.14\% & 16 & 57.14\% & 0 & 0.00\% \\
\hline
parsesUnterminatedTextarea & jsoup & 10 & 10 & 0 & 6 & 60.00\% & 6 & 60.00\% & 0 & 0.00\% \\
\hline
handlesDataOnlyTags & jsoup & 8 & 8 & 0 & 5 & 62.50\% & 5 & 62.50\% & 0 & 0.00\% \\
\hline
outerHtmlGeneration & jsoup & 8 & 8 & 0 & 6 & 75.00\% & 6 & 75.00\% & 0 & 0.00\% \\
\hline
{ \textbf{Mean} } &  & \textbf{20.16} & \textbf{18.06} & \textbf{2.13} & \textbf{15.5} & \textbf{69.42\%} & \textbf{13.73} & \textbf{62.37\%} & \textbf{1.76}  & \textbf{7.04\%}\\
\hline
\end{tabular}
}
\end{center}
\label{tab:results1}
\end{table*}

\begin{figure}
\includegraphics[width=\linewidth, scale=1.0]{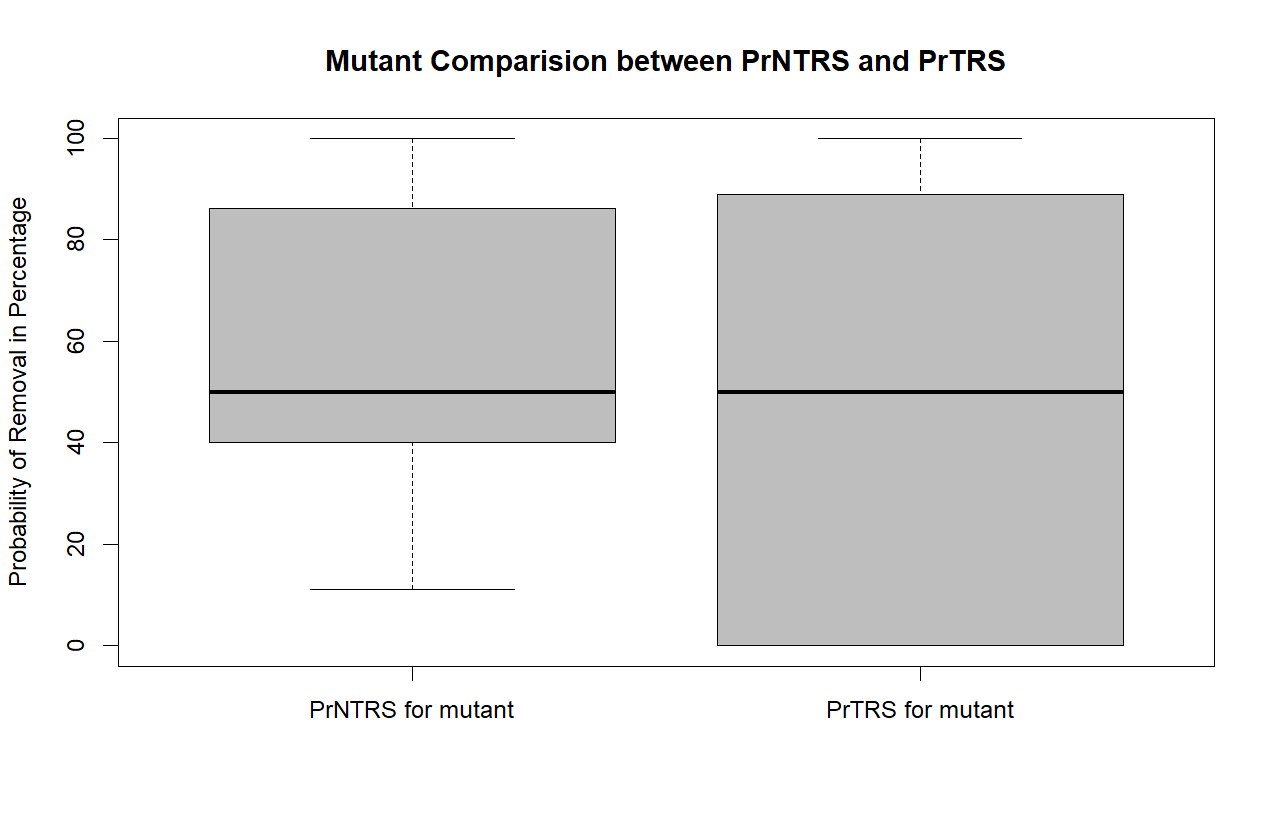}
\caption{Mutants comparison between PrNTRS vs PrTRS }
\label{fig:mutantsPrNTRSvsPrTRS}
\end{figure}

\begin{figure}
\includegraphics[width=\linewidth, scale=1.0]{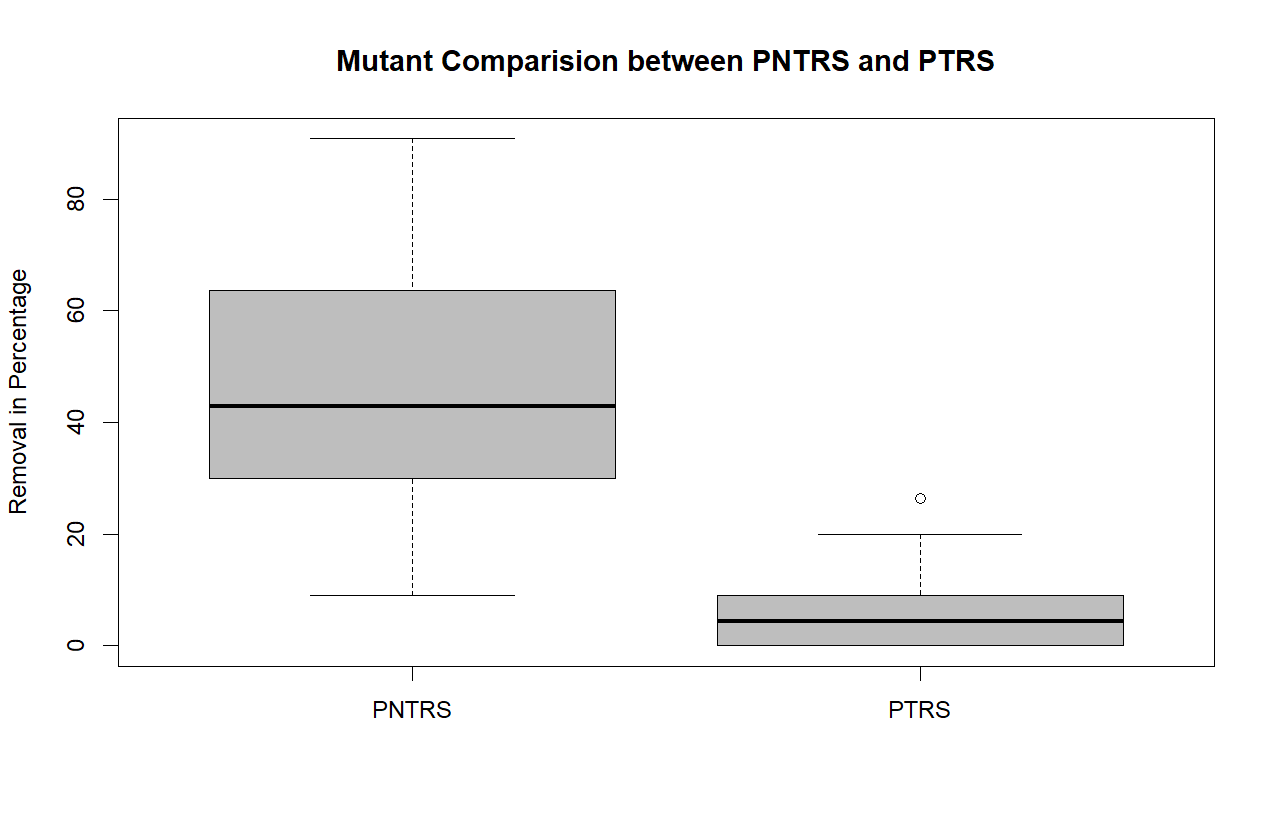}
\caption{Mutants comparison between PNTRS vs PTRS }
\label{fig:mutantsPNTRSvsPTRS}
\end{figure}

\begin{figure}
\includegraphics[width=\linewidth, scale=1.0]{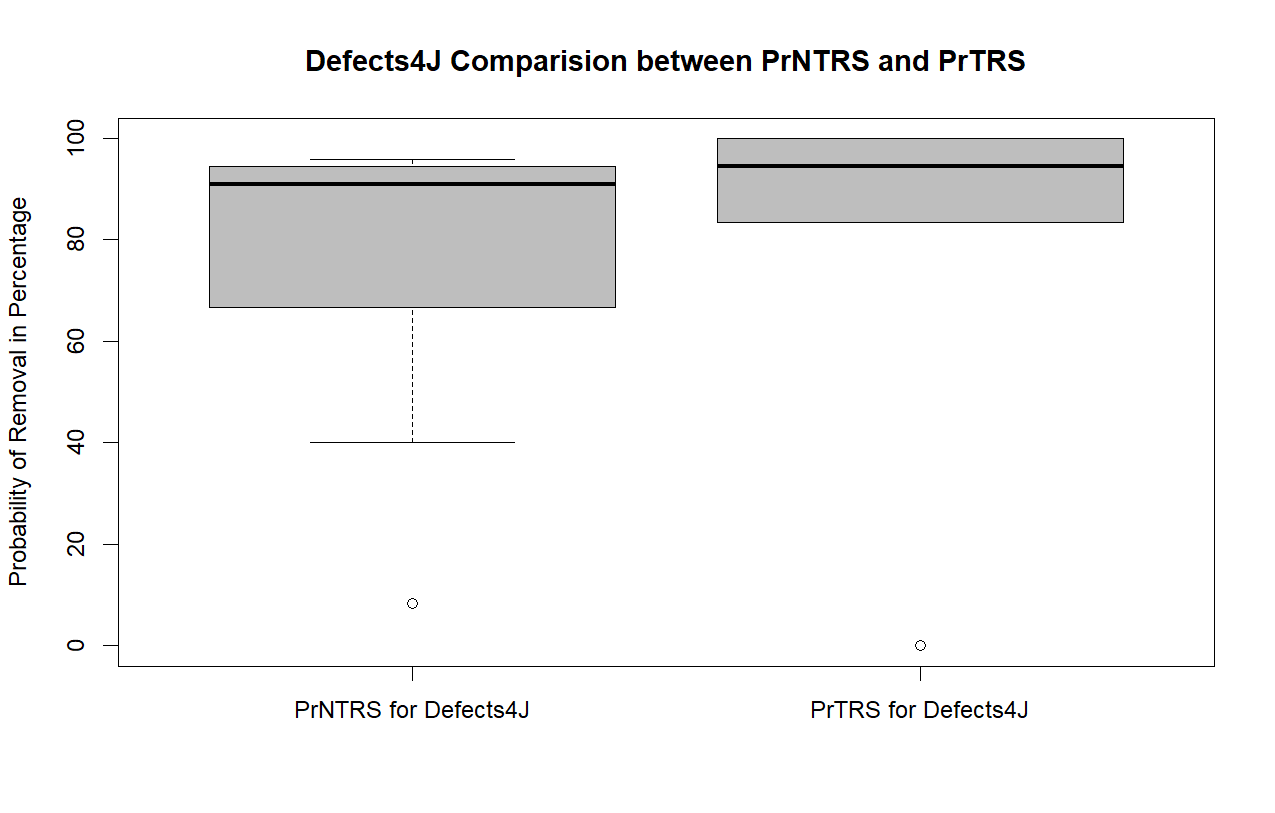}
\caption{\textit{Defects4J} comparison between PrNTRS vs PrTRS }
\label{fig:defect4jPrNTRSvsPrTRS}
\end{figure}

\begin{figure}
\includegraphics[width=\linewidth, scale=1.0]{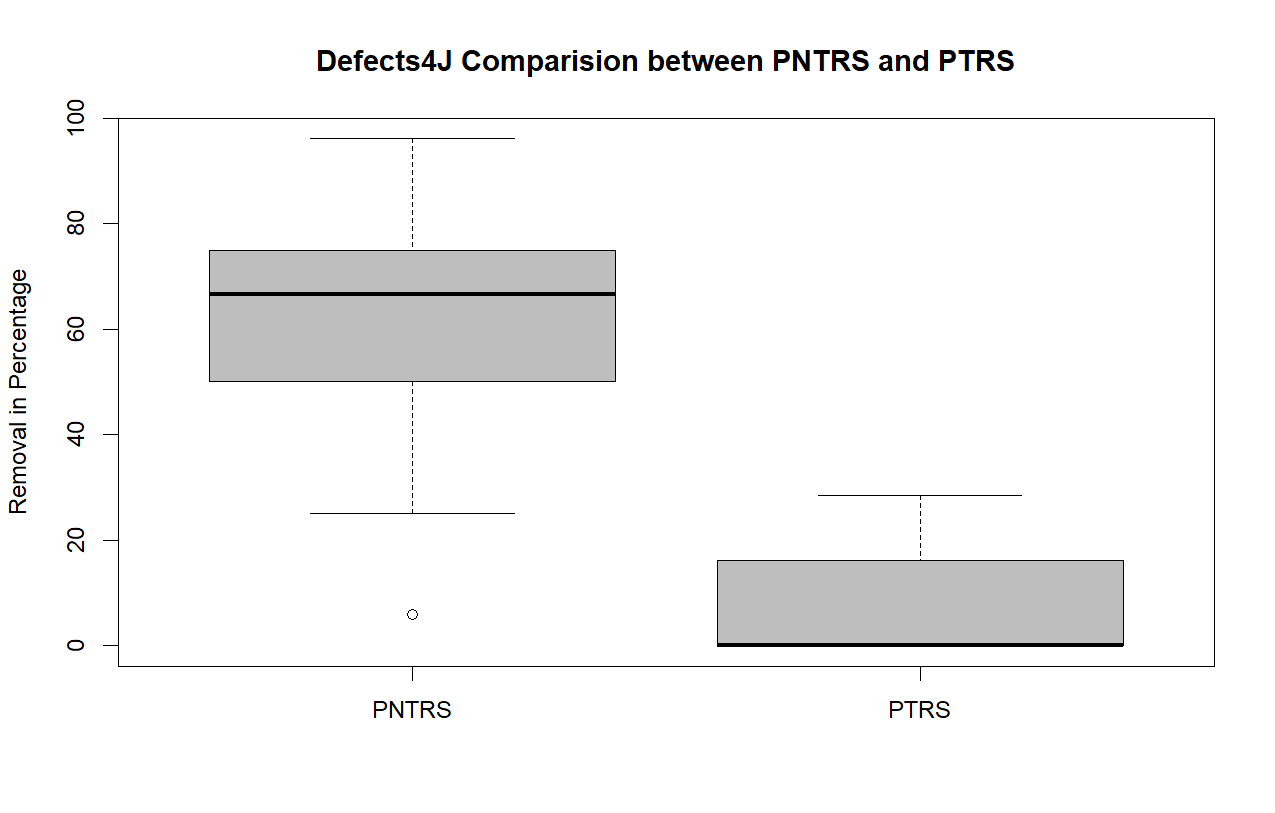}
\caption{\textit{Defects4J} comparison between PNTRS vs PTRS }
\label{fig:defects4jPNTRSvsPTRS}
\end{figure}

\begin{table}
\caption{Mutants Test, PrNTRS vs PrTRS for individual test. Tests that has undefiend PrTRS are not included. }
\begin{center}
{\scriptsize
\begin{tabular}{|l|r| r|}
\hline
Test & PrNTRS & PrTRS   \\
\hline
{testValidatorException} & 50\% & 0.00\% \\
\hline
{testManualBooleanDeprecated} & 86.20\% & 100\% \\
\hline
{checkMissingMethods} & 37.5\% & 0\% \\
\hline
{testManualBoolean} & 86.2\% & 100.00\% \\
\hline
{ExtendedBufferedReaderTest} & 46.15\% & 83.33\% \\
\hline
{testReadLine} & 86.20\% & 80.00\% \\
\hline
{testSerialization} & 85.71\% & 50.00\% \\
\hline
{testJdbcPrinterWithResultSetHeader} & 42.85\% & 0.00\% \\
\hline
{testMySqlNullOutput} & 93.05\% & 88.88\% \\
\hline
{testPrintRecordsWithEmptyVector} & 100\% & 100.00\% \\
\hline
{testLRUBehavior} & 19.23\% & 0.00\% \\
\hline
{testPreparedStatementPooling} & 46.66\% & 50.00\% \\
\hline
{testCallableStatementPooling} & 27.27\% & 0.00\% \\
\hline
{testClosePool} & 50\% & 50\% \\
\hline
{testParameters} & 28.57\% & 100.00\% \\
\hline
{testExcelHeaderCountLessThanData} &71.42\% & 100.00\% \\

\hline
{testManualObject} &100\% & 100.00\% \\
\hline
{testOptionalFinish} &11.12\% & 50\% \\
\hline
{testFileEntryFromPath} & 40\% & 66.66\% \\
\hline
{testExplicitFileEntry} & 50\% & 66.66\% \\
\hline
{testTarArchiveLongNameCreation} & 51.72\% & 50\% \\
\hline
{testContent} & 46.15\% & 50\% \\

\hline
\end{tabular}
}
\end{center}
\label{tab:mutantsPrNTRSvsPrTRS}
\end{table}

\begin{table}
\caption{\textit{Defects4J} Test, PrNTRS vs PrTRS for individual test. Tests that has undefiend PrTRS are not included. }
\begin{center}
{\scriptsize
\begin{tabular}{|l|r| r|}
\hline
Test & PrNTRS & PrTRS   \\
\hline
{testCreateNumber} & 93.93\% & 100\% \\
\hline
{testCalendarTimezoneRespected} & 83.33\% & 100\% \\
\hline
{testLANG807} & 40.00\% & 0\% \\
\hline
{testExceptions} & 94.73\% & 88.88\% \\
\hline
{testConstructorEx7-TypeArray-intArray} & 92.00\% & 83.33\% \\
\hline
{testForOffsetHoursMinutes-int-int} & 95.23\% & 100\% \\
\hline
{testForOffsetHoursMinutes-int-int} & 94.44\% & 83.33\% \\
\hline
{testLocaleIndependence} & 8.30\% & 0.00\% \\
\hline
{testCpioUnarchive} & 90.00\% & 100\% \\
\hline
{testArDelete} & 95.91\% & 100.00\% \\
\hline
{testFinish} & 85.71\% & 100\% \\
\hline
{testRead7ZipMultiVolumeArchiveForStream} & 86.66\% & 83.33\% \\

\hline
{testParseOctalInvalid} & 91.66\% & 87.5\% \\
\hline
{testExcelHeaderCountLessThanData} &66.66\% & 100.00\% \\

\hline
\end{tabular}
}
\end{center}
\label{tab:defects4jPrNTRSvsPrTRS}
\end{table}

\subsection{Process and Measurement}
For each bug and failing test, we apply ReduJavator to generate the \textit{minimal-test}. We compare the failing \textit{oreginal-test} to the failing \textit{minimal-test}. As Christi and Weber did, we measured the following quantities. The definition and example of each quantity is provided in thier work.   

\begin{enumerate}
    \item \textit{absolute-reduction-size }(ARS)
    
    \item  \textit{percentage-reduction-size }(PRS)
    
    \item \textit{absolutte-TreeStmt-reduction-size }(ATRS

    \item \textit{percentage-TreeStmt-reduction-size }(PTRS)
    \item \textit{absolutte-NonTreeStmt-reduction-size }(ANTRS)

    \item \textit{percentage-NonTreeStmt-reduction-size }(PNTRS)

\end{enumerate}

\section{Results}
In this section, we discuss the results for both real-world bugs and the mutants and how the location of a statement affects the reduction. 

For test failures in \textit{Defects4J}, we process 605 statements across 32 classes and 47 test methods across six projects. On average, we processed 121 statements per project. For mutants, we attempt to reduce 627 statements across 30 classes and 30 test methods of open-source projects. Across 30 mutants and failing tests, we process 627 total statements in six different projects. 

The results for \textit{Defects4J} bugs are showin in Table~\ref{tab:results1}. The results for mutatns are shown in Table~\ref{tab:results}. We show total number of statements for each test, the number of \textit{NonTreeStmt}s, the number of \textit{TreeStmts}s, ARS, PRS, ANTRS, PNTRS, ATRS, and PRS.  On average, for mutants we processed 20.9 statements per test that includes 18.86 \textit{NonTreeStmt}s and 2.03 \textit{TreeStmt}s. We reduced on average 12.46 statements or 53.21\% per failing test. For \textit{NonTreeStmt}s, the average reduction was 11.2 or 47.64\%. The same numbers for \textit{TreeStmt}s are 1.23 and 5.38\%. For \textit{Defects4J} failing tests as shown in Table~\ref{tab:results1}, we processed 20.16 statements per test that includes 18.06 \textit{NonTreeStmt}s and 2.13 \textit{TreeStmt}s. We reduce on average 15.5 statements or 69.42\% per failing test. For \textit{NonTreeStmt}s, the average reduction was 13.73 or 63.27\%. The same numbers for \textit{TreeStmt}s are 1.76 and 7.04\%.

The most important entities are the PRS, PNTRS and PTRS. For both the tables Table~\ref{tab:results} and Table~\ref{tab:results1}, half of the PTRS values are 0, and the data is not normally distributed for comparison purposes. We confirm this using the Shapiro-Wilk normality test~\cite{shapiro-test}. Hence, we perform paired Wilcoxon signed rank test on PNTRS vs. PTRS for both \textit{Defects4J} and mutants. For mutants,  $p$-value is  $2.699e-06$ ($p << 0.05$)and V=435~\cite{wilcoxon-test}. We can say that there is a significant difference between PNTRS and PTRS. To demonstrate the difference, we plot both values using a boxplot in Figure~\ref{fig:mutantsPNTRSvsPTRS}.  We can conclude that \textit{NonTreeStmt}s are reduced in large numbers compared to \textit{TreeStmt}s (approximately 9 times). For \textit{Defects4J} $p$-value is  $1.808e-06$ ($p << 0.05$)and V=465. We can say that there is a significant difference between PNTRS and PTRS. To demonstrate the difference we again plot both the values using boxplot in Figrue~\ref{fig:defects4jPNTRSvsPTRS}. We can conclude that \textit{NonTreeStmt}s are reduced in large numbers compared to \textit{TreeStmt}s (approximately 9 times again).

The previous work also calculated the probability of removals by calculating two terms  \textit{PrNTRS} - the probability of removal of \textit{NonTreeStmt}s and (2) \textit{PrTRS} - probability of removal of \textit{TreeStmt}s. For a certain rows in Table~\ref{tab:results} or Table~\ref{tab:results1}, $\#TN$ is 0 and hence \textit{PrTRS} is undefined (divide by 0). We cannot use such tests for evaluation. The results, excluding the tests that have undefined \textit{PrNTRS}, are shown in Table~\ref{tab:mutantsPrNTRSvsPrTRS} for mutants and Table~\ref{tab:defects4jPrNTRSvsPrTRS} for Defects4J. \textit{PrNTRS} and \textit{PrTRS} data is not normal for comparison purposes. We can confirm this using the Shapiro-Wilk test. Hence, to compare \textit{PrNTRS} vs \textit{PrTRS} we again use the paired Wilcoxon signed rank test. For mutants,  $p$-value is  $0.5731$ ($p > 0.05$)and V=109.5. For \textit{Defects4J} $p$-value is  $0.5426$ ($p > 0.05$)and V=42. We can say that for both mutants and \textit{Defects4J}, there is no significant difference between \textit{PrNTRS} vs \textit{PrTRS}. The Comparison between PrNTRS and PrTRS is shown in Figure~\ref{fig:mutantsPrNTRSvsPrTRS} for mutants and in Figure~\ref{fig:defect4jPrNTRSvsPrTRS} for Defects4J.

\subsection{Discussion and Comparison with Previous Work}
In this subsection, we compare our replication results with the previous work in Christi and Weber~\cite{Christi2024OnReducibility}. We both conclude that \textit{NonTreeStmts} are reduced in a large number than \textit{TreeStmts}. However, in their work, the number was 50 times larger. Based on our results, the number is just 9 times larger for both mutants and \textit{Defects4J}. In our case, we cannot conclusively say that non-tree statements have a higher probability of removal than tree-statement. Based on their data, Christi and Weber concluded that non-tree statements have a slightly higher probability of removal than tree statements.

\section{Conclusion}\label{conclusion}
How test reduction algorithms deal with program components (statements in our case) based on the category of the component (tree-statement vs non-tree-statement in our case) needs to be studied further. Christi and Weber studied this and provided the conclusion based on the ReduSharptor tool and C\# subjects. Before accepting any such conclusion, the scientific community needs to conduct replication studies using different subjects and tools, as suggested by Bendrissou et al. In this paper, we conduct one such replication study to demonstrate that our results only partially agree with the conclusion of Christi and Weber. 

As Bendrissou et al. mentioned in their work, a replication study gets less credit than offering a new solution, technique, or ideas, resulting in researchers not attempting to conduct a replication study in Software Engineering. If the scientific community in software engineering starts to pay due respect to replication studies, we may see more replication studies being conducted, ensuring the quality of scientific conclusions.

\bibliographystyle{IEEEtran}

\def\IEEEbibitemsep{0.6pt plus 0.9pt}

\bibliography{IEEEabrv,bibliography}

\end{document}